\documentclass{llncs}
\usepackage{amsmath,amssymb}
\usepackage{type1cm}
\usepackage{algorithm}
\usepackage{algorithmic}
\usepackage[dvipdfmx]{graphicx}
\usepackage{multirow}
\usepackage{booktabs}
\usepackage{comment}
\usepackage{subfig}
\usepackage{adjustbox}
\newcommand{\argmax}{\mathop{\rm arg~max}\limits}

\makeatletter
\def\thickhline{%
  \noalign{\ifnum0=`}\fi\hrule \@height \thickarrayrulewidth \futurelet
   \reserved@a\@xthickhline}
\def\@xthickhline{\ifx\reserved@a\thickhline
               \vskip\doublerulesep
               \vskip-\thickarrayrulewidth
             \fi
      \ifnum0=`{\fi}}
\makeatother

\newlength{\thickarrayrulewidth}
\title{HTML-LSTM: Information Extraction \\from HTML Tables in Web Pages \\using Tree-Structured LSTM}
\author{Kazuki Kawamura\thanks{Now at Sony Group Corporation.}\and Akihiro Yamamoto}
\institute{
Graduate School of Informatics, Kyoto University\\
Yoshida-Honmachi, Sakyo-ku, Kyoto, 606-8501, Japan.\\
\quad
\email{
\hspace{-9pt}Kazuki.Kawamura@sony.com\\
akihiro@i.kyoto-u.ac.jp}
}
\begin{document}
\maketitle
\begin{abstract}
In this paper, we propose a novel method for extracting information from HTML tables with similar contents but with a different structure.
We aim to integrate multiple HTML tables into a single table for retrieval of information containing in various Web pages.
The method is designed by extending tree-structured LSTM, the neural network for tree-structured data, in order to extract information that is both linguistic and structural information of HTML data.
We evaluate the proposed method through experiments using real data published on the WWW.
\end{abstract}

\section{Introduction}
Tables in Web pages are useful for displaying data representing relationships.
We can find them on Web pages showing, for example, syllabus in universities, product information in companies, and flight information in airlines.
Our research aims at integrating tables from various pages but representing the same type of relational data into a single table for retrieval of information.
In this paper, we propose a novel method, called HTML-LSTM, for extracting information from tables in HTML with the same type of contents but with a different structure.

When we browse Web pages, tables representing the same type of relationships look to have similar visual structures but may not be matched completely.
In some tables, every tuple is represented in a row, and in other pages, it is in a column.
The ordering features (or attributes) may be different.
Moreover, they are not always presented as similar HTML source code because different pages are usually designed by different organizations.
The source code may contain noises such as codes for visual decorations and additional information.
Therefore in order to extract and amalgamate relations from tables in different Web pages, we need to unify them in a common set of features (that is, relational schema), as well as in the structure in the level of HTML source codes.
For this purpose, many methods have been proposed~\cite{1683775}, but they work well only for HTML tables having almost the same structure in source codes or for the case where each feature to be extracted clearly differs from each other.
Therefore, extracting and amalgamating relations from tables of similar content but of non-uniform structure is still a major challenge.
We solve this problem by using neural networks developed recently and present our solution as HTML-LSTM.

Some neural networks for extracting effectively features from tree and graph structures have been proposed~\cite{Defferrard2016,Goller1996,Kipf2017,Tai2015}.
The Tree-LSTM~\cite{Tai2015} neural network is a generalization of LSTM to handle tree-structured data, and it is shown that the network effectively works mainly as a feature extractor for parse trees in the field of natural language processing.
Since the source codes of HTML are also parsed into tree structures, we extend Tree-LSTM into HTML-LSTM for extracting features in relational data and tree structure from HTML data simultaneously.

We cannot apply Tree-LSTM to HTML data for the following reason: In parse trees of texts in natural language, linguistic information is given only in the leaves, while the other nodes are given information about the relationship between words.
Therefore, Tree-LSTM transfers information in the direction from the leaves to the root and is often has been applied to tasks such as machine translation and sentiment analysis~\cite{Eriguchi2016,Tai2015}.
On the other hand, when an HTML source code is parsed into a tree, information is attached not only leaves but internal nodes and the root.
In addition, when extracting the features of each element in HTML source codes for tables representing relational data, the path from the root tag \texttt{<table>} to the element is quite essential.
This means that for extracting information from table data, manipulating parsing trees in the direction from the root to leaves as well as in the direction from leaves to the root.
Therefore HTML-LSTM is designed so that information can be transferred in both directions.

In applying HTML-LSTM to information extraction from real HTML data, we first extract the substructure of a table in the data and convert it into a tree.
Next, we extract features from the obtained tree structure using HTML-LSTM and classify the features to be extracted for each node.
Finally, we integrate the extracted information into a new table.
We also introduce a novel data augmentation method for HTML data in order to improve the generalization performance of information extraction.

We evaluate and confirm the effectiveness of the HTML-LSTM method for integrating HTML data formats by applying the method explained above to tables of preschools of local governments and tables of syllabuses published by universities, which are published on the Web.
As the results, we succeeded in extracting the required information with an accuracy of an $F_1$-measure of 0.96 for the data of preschools and an $F_1$-measure of 0.86 for the syllabus data.
Furthermore, our experimental results show that HTML-LSTM outperforms Tree-LSTM.

This paper is organized as follows.
In Section 2, we describe previous methods for extracting information from Web pages.
In Section 3, we introduce the architecture of our proposed method, HTML-LSTM, and how to use HTML-LSTM for information extraction.
In Section 4, we summarize the results of experiments using HTML data on the Web.
Finally, in Section 5, we provide our conclusion.

\section{Related Work}
Extracting information from documents and websites and organizing it into a user-friendly form is called information extraction and is widely studied.
The research field originates with the Message Understanding Conference (MUC)~\cite{Grishman1996,Sundheim1992}, which started in the 1980s.
At this conference, every year, a competition is held to extract information from newspapers on, for example, terrorist activities, product development, personnel changes, corporate mergers, rocket launches, and participants competed for some scores evaluating their technique for information extraction.

In the early years of the research area, rule-based methods were widely used~\cite{Cunningham2002,Shaalan2001}, where rules are defined based on features such as the representation of the characters of the tokens in the sentence, the notation of the tokens (uppercase, lowercase, mixed case of uppercase and lowercase, spaces, punctuation, \textit{etc.}), and the parts of speech of the tokens.
Such rule-based methods require experts who manually create rules depending on the types of objects they want to extract.
Since it is very time-consuming, algorithms have been developed to automatically create rules using labeled data~\cite{Aitken2002,Ciravegna2001,Soderland1999}.
In recent years, statistical methods have also been used in order to treat documents which may have many noises.
Example of methods are Support Vector Machine (SVM)~\cite{Koichi2002}, Hidden Markov Model (HMM)~\cite{Seymore1999}, Maximum Entropy Markov Model (MEMM)~\cite{McCallum2000}, Conditional Markov Model (CMM)~\cite{Malouf2002}, and Conditional Random Fields (CRF)~\cite{Peng2006}.


Our key idea is to introduce natural language processing methods for information extraction and simultaneously handle structural and linguistic information.
Every Web page is written as a source code in HTML, with a clear tree structure after parsing it.
The method to extract a specific part from a Web page using the structural information is called Web wrapper~\cite{Kushmerick2000}.
Some of the methods extract information by regarding a specific part in a Web page as data in a tree structure~\cite{Kashima2002,Muslea1999}.
These methods work for Web pages of almost similar structure, and it is difficult to apply them to pages whose structure is completely different, but the meaning of them is the same.
This situation often appears in tables representing relational data.

Other types of methods are treating linguistic features of Web pages based on natural language processing, in other words, treating the meaning of the texts on each page.
These methods have the disadvantage that they cannot capture the structure of pages.
However, natural processing has greatly advanced thanks to the introduction of greatly improved neural network techniques.
Some researchers propose new types of neural networks which treat the parsing tree of texts in natural languages.
This motivates us to apply such neural networks to extracting information taking into account structure and meaning simultaneously.

\newpage
\section{HTML-LSTM}

\begin{figure}[t]
	\centering
	\includegraphics[width=1.0\textwidth]{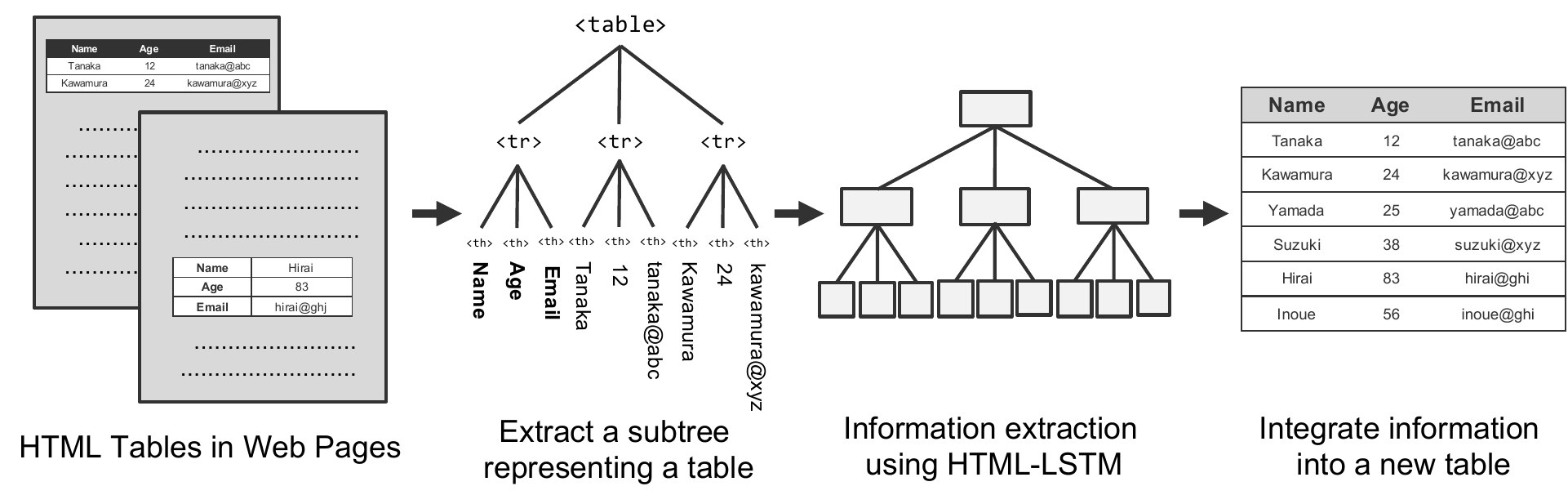}
	\caption{The {\bf HTML-LSTM} framework for information extraction and integration from HTML tables in web pages}
  \label{fig:overview}
\end{figure}

The overview of our proposed method is shown in Fig.~\ref{fig:overview}.
First, we extract the substructure of the table from the entire HTML data.
Since Table tags (\texttt{<table>} \texttt{</table>}) are usually used to represent tables in HTML, we can extract the table by focusing on the region surrounded by the table. tags.
Next, we convert the HTML data into a tree structure called DOM tree so that HTML-LSTM can take it as its input.
Then, the obtained tree structure data is input to HTML-LSTM for feature extraction, and the obtained features of nodes are used to classify which attribute values each node belongs to.
Finally, we pick up the nodes' information classified into the extraction target's attribute values and integrate the information in a new single table.

\subsection{Extracting Information}
\begin{figure}[t]
	\centering
	\includegraphics[width=1.0\textwidth]{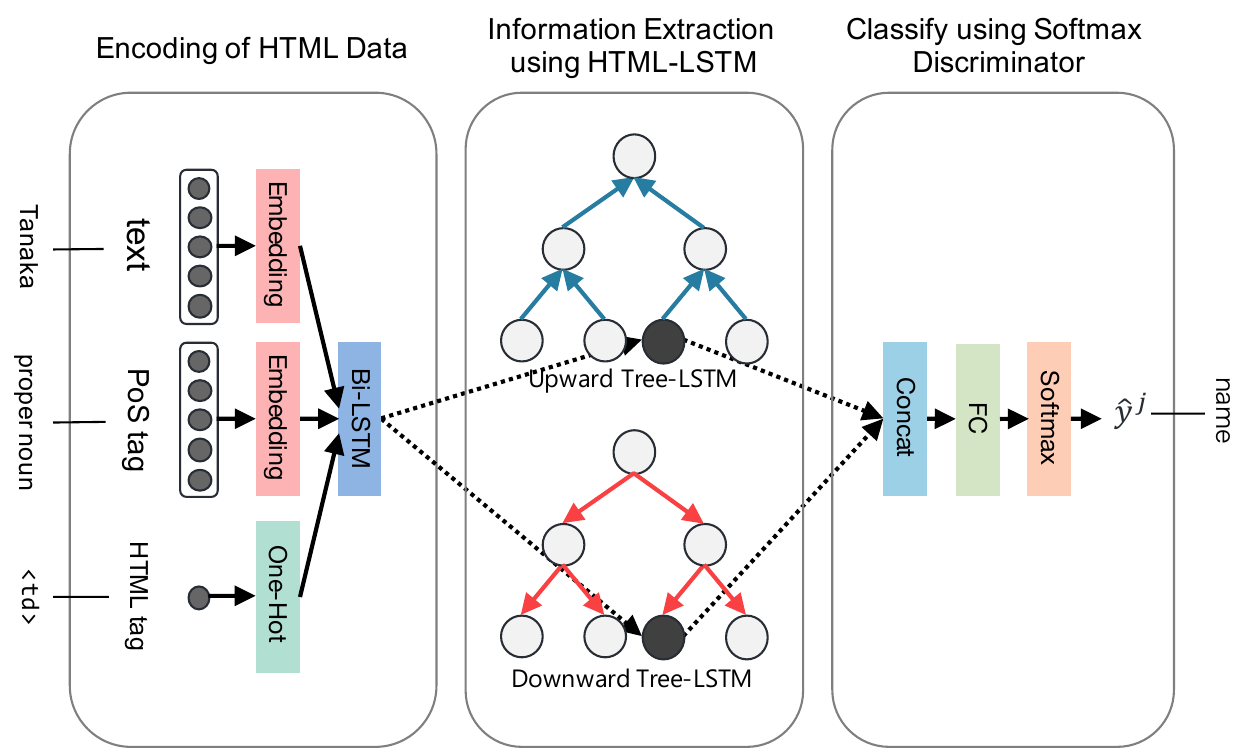}
	\caption{The workflow of information extraction using HTML-LSTM}
  \label{fig:information_extraction}
\end{figure}

In this subsection, we explain the details of HTML-TLSM and the extraction of information from HTML data using it.
The workflow of information extraction is shown in Fig.~\ref{fig:information_extraction}.
First, each element of the HTML data is encoded using Bi-LSTM (Bidirectional LSTM)~\cite{Hochreiter1997a,Schuster1997a} in order to obtain the language representation of each element.
Next, HTML-LSTM is applied to obtain the features of the HTML data, considering the relationship between the positions of the elements in the parsed tree.
In order to use the information of the whole tree structure of HTML data effectively, HTML-LSTM extends Tree-LSTM, in which the information flows only from leaf to root, to enable the flow of information from root to leaf as well as from leaf to root.
Finally, the features of each node obtained by the HTML-LSTM are passed through the fully connected layer, and the softmax classifier is applied to determining which attribute value each node is classified as.

\subsubsection{Encoding of HTML Data:}
\begin{figure}[t]
	\centering
	\includegraphics[width=1.0\textwidth]{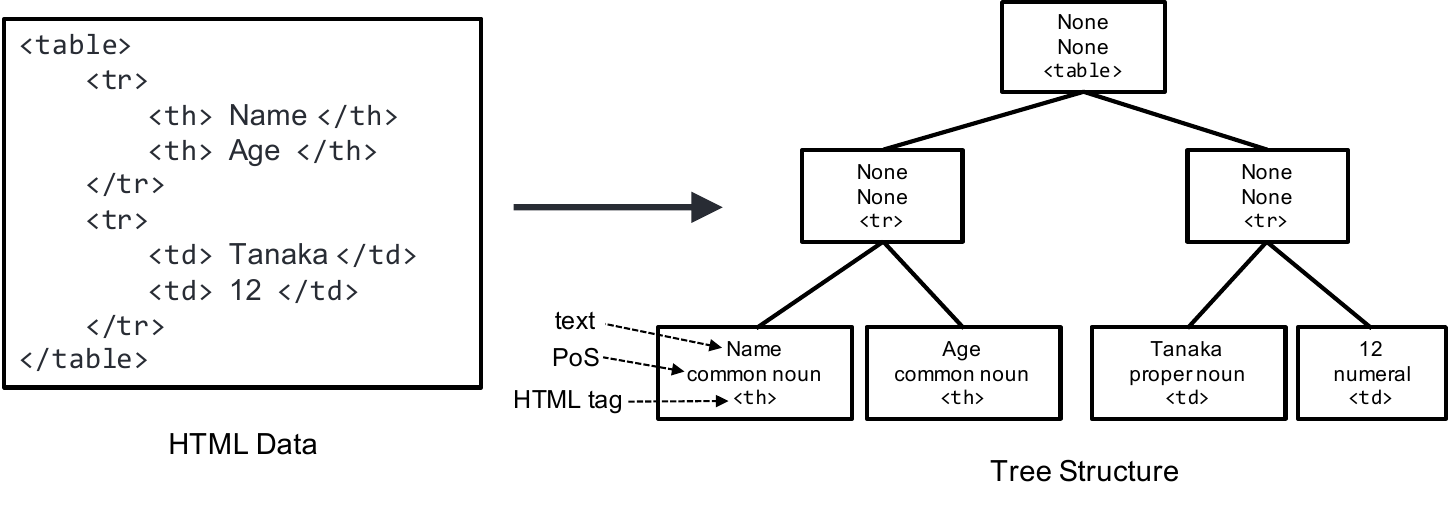}
	\caption{Example of converting HTML data to a tree structure}
  \label{fig:html2tree}
\end{figure}

The DOM tree that is fed into HTML-LSTM is obtained by parsing the HTML source code.
In general, when parsing HTML data to a tree, the values of the nodes in the tree structure are HTML tags.
In our method, in order to extract information by effective use of the linguistic and structural information of HTML, each node of the tree has three types of values: the HTML tag, the text between the start and end tags, and the PoS (part-of-speech) tags of the text, as shown in Fig.~\ref{fig:html2tree}.
The extracted text is treated as a sequence of words.
The PoS tag is the sequence of parts-of-speech data corresponding to the sequence of words of the text.
If the attribute names to be extracted differ from a Web page to a Web page, a dictionary is used to unify the attribute names.
Finally, the obtained tree is converted into a binary tree because HTML-LSTM accepts only binary trees as input.

After converting the HTML data into a binary tree, the text, the sequence of PoS tags, and the HTML tag in each node are converted using a neural network to a representation for input to the HTML-LSTM.
In particular, we combine the one-hot encoding of the tag $t^j$ of a node $j$ and the text $w^j$ and PoS tags $p^j$  converted by the embedding matrices $E_{\rm content}$ and $E_{\rm pos}$, and feed them to Bi-LSTM:
\begin{align*}
e^j_t &= \bigl[{\rm onehot}(t^j) \| E_{\rm content}(w^j_t) \| E_{\rm pos}(p^j_{t})\bigr
], \\
\overrightarrow{h^j_{t}} &= \overrightarrow{\rm LSTM}\bigl(e^j_t, \overrightarrow{h^j_{t-1}}\bigr), \\
\overleftarrow{h^j_{t}} &= \overleftarrow{\rm LSTM}\bigl(e^j_t, \overleftarrow{h^j_{t-1}}\bigr),
\end{align*}
where ${\rm onehot}$ is a function that converts a tensor to one-hot a representation and $\|$ is the concatenation of two tensors.
The function $\overrightarrow{\rm LSTM}$ is the forward LSTM and $\overleftarrow{\rm LSTM}$ is the backward LSTM of the Bi-LSTM.
The outputs at the last time $T$ of the forward and backward LSTMs are combined to obtain the representation $x^j$ = $\bigl[\overrightarrow{h_T} \| \overleftarrow{h_T}\bigr]$ for each node.

\subsubsection{HTML-LSTM:}
\begin{figure}[p]
  \centering
  \subfloat[Upward Tree-LSTM]{
    \includegraphics[width=1.\textwidth]{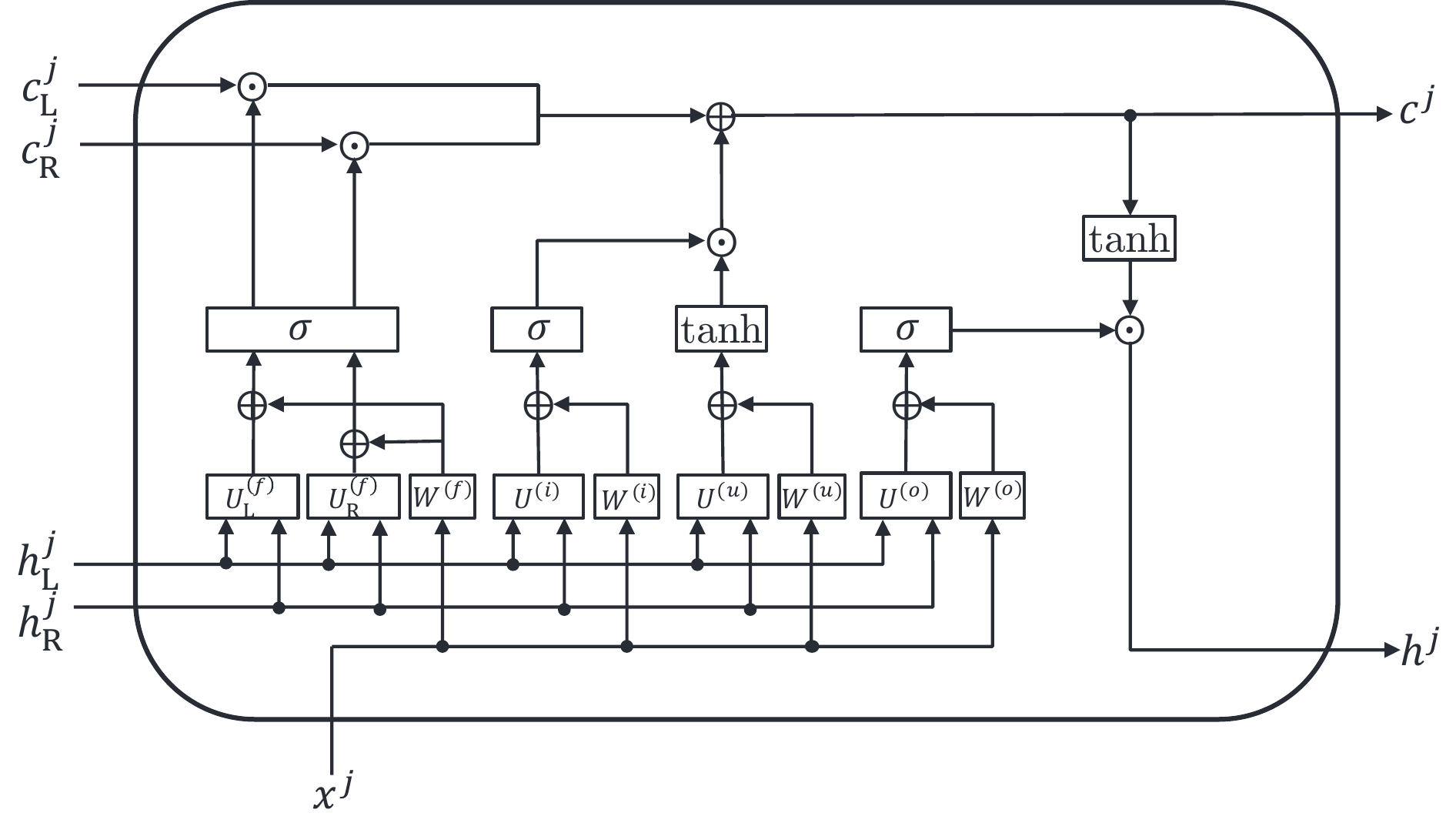}
    \vspace{-5mm}
    \label{fig:upward-tree-lstm}
  } \hfil \vspace{10mm}
  \subfloat[Downward Tree-LSTM]{
    \includegraphics[width=1.\textwidth]{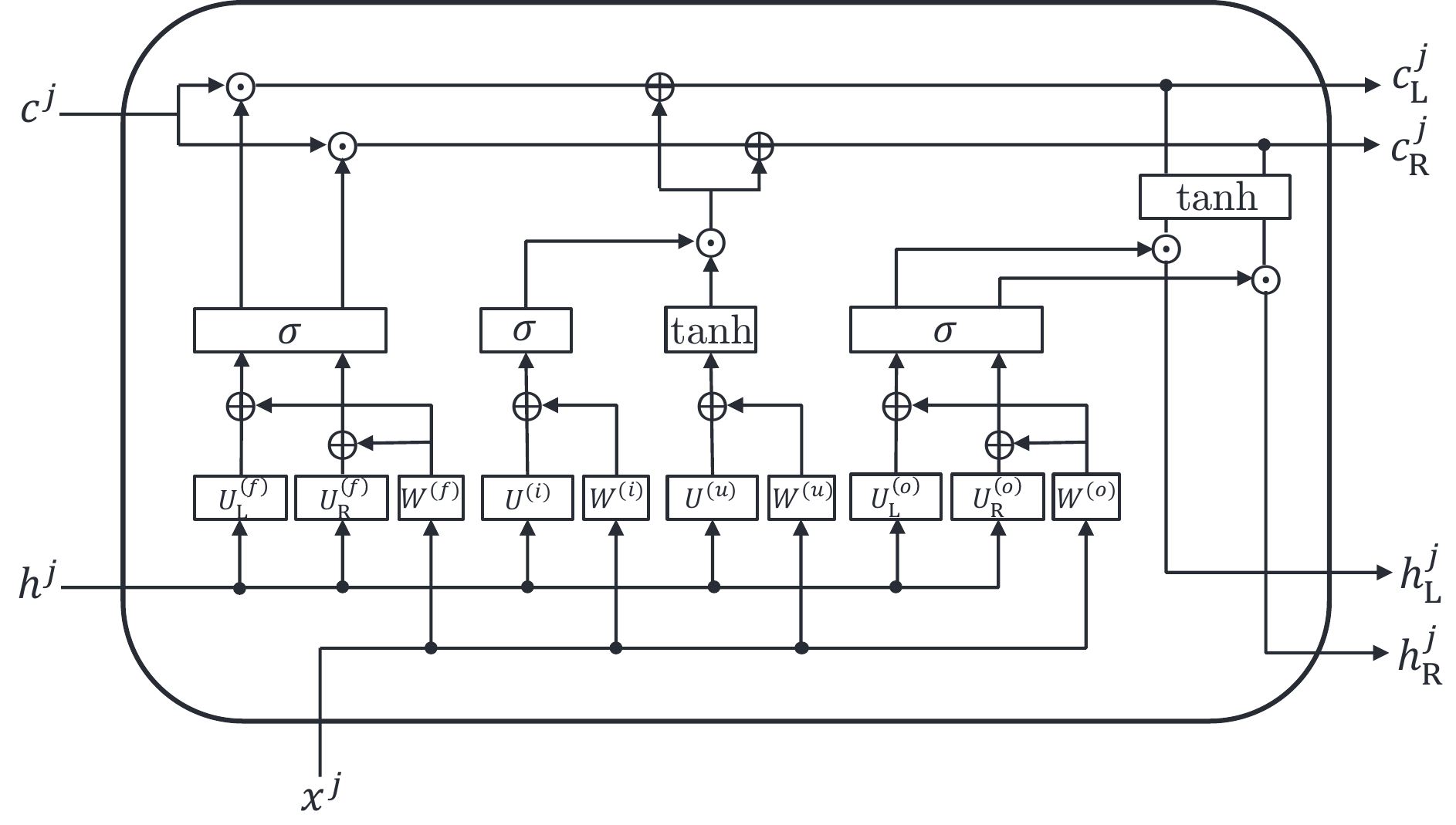}
    \vspace{-5mm}
    \label{fig:downward-tree-lstm}
  } \vspace{5mm}
  \caption{HTML-LSTM architecture}
  \label{fig:html-lstm}
\end{figure}

HTML-LSTM is composed of Upward Tree-LSTM, in which information flows from leaves to roots, and Downward Tree-LSTM, in which information is transmitted from roots to leaves.
Upward Tree-LSTM uses Binary Tree-LSTM~\cite{Tai2015}. The model is expressed as follows:
\begin{align*}
	i^{j} &=\sigma \Bigl(W^{\scalebox{0.60}{$(i)$}} x^{j} + \sum_{k\in\{{\rm L}, {\rm R}\}}{U_{k}^{\scalebox{0.60}{$(i)$}} h^j_k} + b^{\scalebox{0.60}{$(i)$}}\Bigr), \\
	f^j_{\scalebox{0.55}{$\left[{\rm L},{\rm R}\right]$}} &=\sigma \Bigl(W^{\scalebox{0.60}{$(f)$}} x^{j} + \sum_{k\in\{{\rm L}, {\rm R}\}}{U_{\scalebox{0.55}{$\left[{\rm L},{\rm R}\right]$}k}^{\scalebox{0.60}{$(f)$}} h^j_k} + b^{\scalebox{0.60}{$(f)$}}\Bigr), \\
	o^{j} &=\sigma \Bigl(W^{\scalebox{0.60}{$(o)$}} x^{j} + \sum_{k\in\{{\rm L}, {\rm R}\}}{U_{k}^{\scalebox{0.60}{$(o)$}} h^j_k} + b^{\scalebox{0.60}{$(o)$}}\Bigr), \\
	u^{j} &=\tanh \Bigl(W^{\scalebox{0.60}{$(u)$}} x^{j} + \sum_{k\in\{{\rm L}, {\rm R}\}}{U_{k}^{\scalebox{0.60}{$(u)$}} h^j_k} + b^{\scalebox{0.60}{$(u)$}}\Bigr), \\
	c^{j} &=i^{j} \odot u^{j} + \sum_{k\in\{{\rm L}, {\rm R}\}}{f^j_k \odot c^j_k}, \\
	h^{j} &=o^{j} \odot \tanh \left(c^{j}\right).
\end{align*}
Upward Tree-LSTM has a forget gate $f^j$, an input gate $i^j$, an output gate $o^j$, a memory cell $c^j$, and a hidden state $h^j$, just like a simple LSTM.
In the expressions, $\sigma$ denotes the sigmoid function and $\odot$ denotes the element-wise product.
Both of the parameters $W$ and $U$ are weights, and $b$ is the bias.
All of these parameters are learnable.
As shown in Fig.~\ref{fig:upward-tree-lstm}, Upward Tree-LSTM is a mechanism that takes two inputs and gives one output.
In this case, the forget gate uses its own parameters $U_{\rm L}$ and $U_{\rm R}$ to select the left and right children's information $c^j_{\rm L}, c^j_{\rm R}$.

On the other hand, Downward Tree-LSTM, in which information flows from roots to leaves, is as follows:
\begin{align*}
	i^{j} &=\sigma \left(W^{\scalebox{0.60}{$(i)$}} x^{j} + U^{\scalebox{0.60}{$(i)$}} h^{j}  + b^{\scalebox{0.60}{$(i)$}}\right), \\
	f^j_{\scalebox{0.55}{$\left[{\rm L},{\rm R}\right]$}} &=\sigma \left(W^{\scalebox{0.60}{$(f)$}} x^{j} + U_{\scalebox{0.55}{$\left[{\rm L},{\rm R}\right]$}}^{\scalebox{0.60}{$(f)$}} h^{j} + b^{\scalebox{0.60}{$(f)$}}\right), \\
	o^j_{\scalebox{0.55}{$\left[{\rm L},{\rm R}\right]$}} &=\sigma \left(W^{\scalebox{0.60}{$(o)$}} x^{j} + U_{\scalebox{0.55}{$\left[{\rm L},{\rm R}\right]$}}^{\scalebox{0.60}{$(o)$}} h^{j} + b^{\scalebox{0.60}{$(o)$}}\right), \\
	u^{j} &=\tanh \left(W^{\scalebox{0.60}{$(u)$}} x^{j} + U^{\scalebox{0.60}{$(u)$}} h^{j} + b^{\scalebox{0.60}{$(u)$}}\right), \\
	c^j_{\scalebox{0.55}{$\left[{\rm L},{\rm R}\right]$}} &=i^{j} \odot u^{j} + \sum_{k\in\{{\rm L}, {\rm R}\}}{f^j_k \odot c^{j}}, \\
	h^j_{\scalebox{0.55}{$\left[{\rm L},{\rm R}\right]$}} &= o^j_{\scalebox{0.55}{$\left[{\rm L},{\rm R}\right]$}} \odot \tanh(c^j_{\scalebox{0.55}{$\left[{\rm L},{\rm R}\right]$}}).
\end{align*}
As shown in Fig.~\ref{fig:downward-tree-lstm}, this mechanism takes one input and gives two outputs. In this case, the forgetting gate generates two outputs by operating on the input $c^j$ with different parameters $U_L$ and $U_R$, so that the model can choose information to transmit to the left and right children.

Finally, we combine the Upward Tree-LSTM hidden state $h^j_{\uparrow}$ and the Downward Tree-LSTM hidden states $h^j_{l\downarrow}$, $h^j_{jr\downarrow}$ to obtain $h^j$ = $\left[h_{j\uparrow} \\| h_{jl\downarrow} \| h_{jr\downarrow}\right]$ as the feature of each node.
A softmax classifier predicts the label $\hat{y}^{j}$ from among the $N$ classes,
\begin{align*}
	p &= \operatorname{softmax}\left(W^{(s)} h^{j}+b^{(s)}\right), \\
	\hat{y}^{j} &= \argmax_{i\in{\{1,\ldots,N\}}}{p_i},
\end{align*}
where $W^{(s)}, b^{(s)}$ are the learnable parameters.
The set of labels consists of the set of attributes to be extracted and a special label \textit{Other} that does not belong to any of the attributes to be extracted.
For example, when extracting the attributes \textit{Name} and \textit{Age} from the table shown in Fig.~\ref{fig:overview} (left), there are three types of classes: \textit{Name}, \textit{Age}, and \textit{Other}.
The attribute values ``Hirai'' and ``8'' in the table belong to the class of \textit{Name} and \textit{Age}, respectively, while the attribute value ``hirai@ghi'' and the attribute names ``Name'', ``Age'', and ``Email'' belong to the class of \textit{Other}.
We classify all the nodes in the HTML tree to determine what attribute each node is (or is not included in any of the attributes to be extracted).

Every HTML source code may contain much information that is not required to be extracted and noise for decoration.
In treating real data, most nodes are not the target of extraction, and therefore the trees as the inputs of HTML-LSTM tend to be imbalanced.
In order to treat such trees, we use Focal Loss~\cite{Lin2020} which is an extension of Cross-entropy Loss to deal with class imbalance.
Focal Loss is defined as follows with the correct label $N$ and one-hot vector $t$:
\begin{align*}
	\mathcal{L}_{\rm focal} = -\alpha_{i}\sum_{i=1}^N\left(1-p_{i}\right)^{\gamma} t_{i} \log \left(p_{i}\right),
\end{align*}
where $\alpha_{i}$ is the frequency inverse of each class and $\gamma$ is the hyperparameter.
\newpage \noindent
Also, in order to improve scores of the model's recall and precision in a well-balanced way, F1 loss is used jointly.
The F1 Loss is given by $1-F_1$, where ${F_1}$ is the average of the $F_1$ measures of each class.
This is denoted as $\mathcal{L}_{\rm f1}$, and the final loss function $\mathcal{L}$ is given as
\begin{align*}
  \mathcal{L} = \mathcal{L}_{\rm focal} + \mathcal{L}_{\rm f1}.
\end{align*}

Furthermore, the table data on the Web is equivalent to the original data even if the order of the rows and columns of the table is changed.
Therefore, we introduce a data augmentation technique that randomly changes the order of rows and columns, thereby increasing the number of HTML data used for training.

\subsection{Integrating Information}\label{sec:information_integration}
\begin{figure}[hbt]
	\centering
	\includegraphics[width=1.0\textwidth]{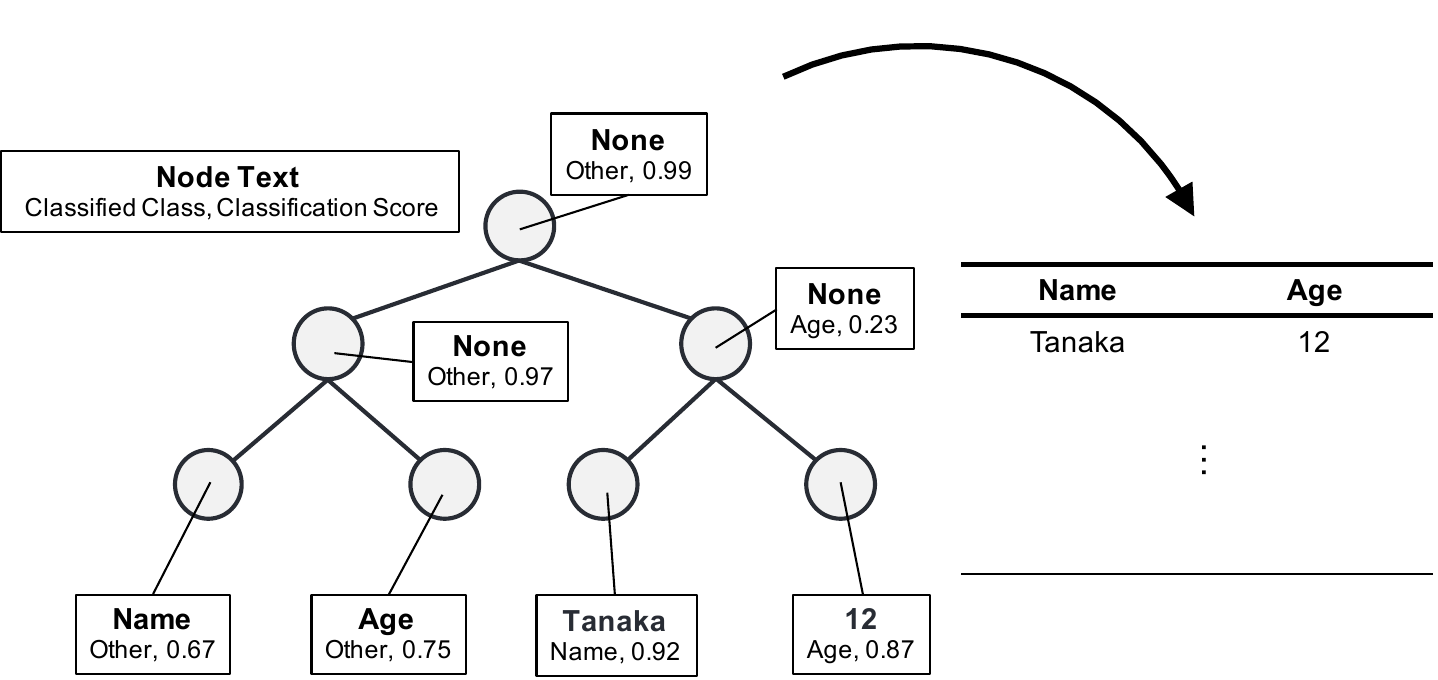}
	\caption{Example of information integration}
  \label{fig:information_integration}
\end{figure}

After classifying the class of each node in the HTML tree using HTML-LSTM, we extract the required information from the tree and integrate it into a new table.
For each class (attribute), the node with the highest classification score in the HTML tree is selected, and the text of that node is extracted and put into the table.
Here, the classification score is the maximum value of the output of the softmax classifier for each class, \textit{i.e.}, $\underset{i}{\max}~p_i$.
The left side of Fig.~\ref{fig:information_integration} shows an example of the tree after classifying the class of each node.
Each node contains the text of the original element (top row in the box), the class with the highest classification score (bottom row left in the box), and the classification score (bottom row right in the box).
For example, to extract the information of \textit{Name} class from this tree, extract the text of the node elements classified as \textit{Name} class.
In this case, the only node classified in the \textit{Name} class is ``Tanaka'', so we extract ``Tanaka'' as the \textit{Name} information of this HTML tree and put it in the table.
In the same way, when we extract the \textit{Age} class information, we find that there are two nodes classified as \textit{Age} class: ``12'' and ``None''.
Comparing the classification scores, the classification score of ``12'' is $0.87$, and ``None'' is $0.23$.
Since the score of ``12'' is higher, we extract ``12'' as the \textit{Age} information of this HTML tree and put it in the table.
However, if multiple values should be classified into a certain class, there will be omissions in the extraction.
Therefore, when it is expected that there are multiple values classified into a certain class in the web table, a threshold value is set, and all the texts of the nodes that exceed the threshold value are extracted.
For example, suppose that in a given HTML tree, there are three nodes classified into the \textit{Name} class, ``Tanaka'', ``Suzuki'', and ``Apple'' with classification scores of $0.97$, $0.89$, and $0.23$, respectively.
If the threshold is set to $0.5$, then ``Tanaka'' and ``Suzuki'' will be extracted as the \textit{Name} class information for this tree.

\subsection{Implementation Details}
The dimensions of the hidden layer of the model are $128$ and $5$ for the embedding layer of the text and part-of-speech tags, respectively, in the information embedding part of the HTML data.
The dimensions of the hidden layer of the HTML-LSTM are $64$, and the dimensions of the linear layer used for classification are $64$.

We use Adam~\cite{Kingma2015} as the optimization algorithm, with a minibach size of $128$.
The learning rate starts from $10^{-2}$ and is divided $2$ every $15$ epochs, and the model is trained for $50$ epochs.
The parameters of Adam are $\alpha$ of $10^{-2}$, $\beta_1$ of $0.9$ and $\beta_2$ of $0.999$.
We use Dropout~\cite{Hinton2014} with a probability of $0.5$ is used to prevent overfitting.

\section{Experiments}
We evaluate our method on tables of preschools published by local governments and tables of syllabus published by universities.
These data are published on the Web, and the information is presented using table structures represented by HTML.
For evaluation, we use Recall, Precision, and their harmonic mean, $F_1$ measure.
$F_1$ measure and Precision and Recall are defined by
\begin{gather*}
  F_{1}=2 \cdot \frac{\text{Precision} \cdot \text{Recall}}{\text{Precision}+\text{Recall}}, \\
  \text{Precision} =\frac{TP}{TP+FP}, \quad \text{Recall} =\frac{TP}{TP+FN},
\end{gather*}
where $TP$, $TN$, $FP$, and $FN$ are the true positives,
true negatives, false positives, false negatives, respectively.

\begin{table}[t]
	  \begin{center}
	    \caption{Information extraction result for the preschool data}
	    \label{tab:result_extractinon_nursery}
	    \begin{tabular}{lccc} \thickhline
			 Attribute & Precision & Recall & $F_1$ measure \\ \hline
	     name & 0.94 & 1 & 0.97 \\
	     address & 0.92 & 1 & 0.96 \\
	     telephone number & 0.87 & 1 & 0.92 \\
	     other & 1 & 0.98 & 0.99 \\
	     mean & 0.93 & 0.99 & 0.96 \\ \thickhline
	    \end{tabular}
	  \end{center}
\end{table}

\subsection{Experiments on Preschool Data}
Many local governments publish a list of preschools in their localities in a table on the Web.
Those pages have common information, such as the preschools' name, address, and phone number, but each page has a different HTML structure.
In this experiment, we extracted and integrated the information of \textit{name}, \textit{address}, and \textit{phone number} from these pages.
We collected a total of 47 HTML data from 47 local governments that contain information on preschools for the experiment.
The HTML data were converted into a tree with a text, PoS tags, and HTML tags at each node.
The obtained ordered trees had 22--249 nodes (107 nodes on average) and contained 16 types of PoS tags.
Since the data collected for the experiment was in Japanese, the word boundaries are not obvious.
Therefore, the series of text and PoS tags were obtained by morphological analysis using Janome\footnote{https://mocobeta.github.io/janome/}.
We also unified attribute names that have the same meaning but different notation, and labeled each node manually.
The classes were four types of labels: \textit{name, address, phone number}, and \textit{other}.

Table~\ref{tab:result_extractinon_nursery} shows the results of information extraction for each attribute in the case of $10$ fold cross-validation of preschool data.
Table~\ref{tab:result_integration_nursery} shows the integrated table of the information extracted from the preschool data.
We can see that our model does good work on information extraction and integration from these results.

\begin{table}[t]
  \begin{center}
    \caption{Information integration result for the preschool data}
    \label{tab:result_integration_nursery}
      \scalebox{1.}{
      \begin{tabular}{|c|c|c|} \thickhline
       Name & Address & Telephone number \\ \hline
       \begin{tabular}{c}
       Tourin Kindergarten
       \end{tabular} &
       \begin{tabular}{c}
       43 Shichiku Takanawacho Kita-ku
       \end{tabular} & 492-4717 \\
       \begin{tabular}{c}
       Kamo Nursery Room
       \end{tabular} &
       \begin{tabular}{c}
       59 Kamigamo Ikedonocho Kita-ku
       \end{tabular} & 585-5958 \\
       \begin{tabular}{c}
       Nonohana Preschool
       \end{tabular} &
       \begin{tabular}{c}
       37 Koyama Nishionocho Kita-ku
       \end{tabular} & 354-6927 \\
       \begin{tabular}{c}
       Nisihuji Preschool
       \end{tabular} &
       \begin{tabular}{c}
       1584-1 Nishifujicho
       \end{tabular} & 0848-55-6920 \\
       \begin{tabular}{c}
       Mitsugi Chuo Preschool
       \end{tabular} &
       \begin{tabular}{c}
       94 Mitsugicho Hanajiri
       \end{tabular} & 0848-76-0044 \\ \thickhline
     \end{tabular}}
  \end{center}
\end{table}

\subsection{Experiments on Syllabus Data}
The syllabus is the data which shows the contents of lectures in universities and is published on the Web by many universities, mainly using a table format.
We collected the syllabus from 22 different universities on the Web and used the HTML data of 20,257 pages for the experiment.
The syllabus data was converted into a tree structure in the same way as the data of preschools, and the attribute names with different notations were unified and labeled.
The obtained ordered tree has 19 to 1,591 nodes (109 on average) and contains 25 kinds of tags.
Since some of the obtained trees contain much noise other than necessary information, we clipped the nodes after the 100th node in the post-order in the syllabus data.
The extracted attributes are \textit{course title, instructor name, target student, target Year, year/term, day/period}, and \textit{number of credits}.
Therefore, there are eight types of labels in the syllabus data, including \textit{other} in addition to these seven classes.

Table~\ref{tab:result_extractinon_syllabus} shows the results of information extraction for each attribute in the case of $5$ fold cross-validation of syllabus data.
In each split, 18 (or 17) of the 22 universities were used for training, and 4 (or 5) were used for testing.
Table~\ref{tab:result_integration_syllabus} shows the integrated table of the information extracted from the syllabus data.
The $F_1$ measure of the information extraction in the syllabus data is less than that in the preschool data.
We believe this is because more attributes are extracted than in the preschool data, and more noise is included in the syllabus data.
The result of the information integration shows that information can be extracted even in the blank areas (\textit{i.e.}, areas that originally had no information).
This shows that our model can use not only linguistic information but also structural information.

\begin{table}[t]
  \begin{center}
    \caption{Information extraction result for the syllabus data}
    \label{tab:result_extractinon_syllabus}
    \begin{tabular}{lccc} \thickhline
     Attribute & Precision & Recall & $F_1$ measure \\ \hline
     course title & 0.76 & 0.82 & 0.77 \\
     instructor name & 0.81 & 0.79 & 0.80 \\
     target student & 0.90 & 0.76 & 0.82 \\
     target year & 0.94 & 0.75 & 0.83 \\
     year/term & 0.97 & 0.83 & 0.89 \\
     day/period & 0.89 & 0.87 & 0.88 \\
     number of credits & 0.83 & 0.94 & 0.88 \\
     other & 0.99 & 0.99 & 0.99 \\
     mean & 0.89 & 0.84 & 0.86 \\ \thickhline
    \end{tabular}
  \end{center}
\end{table}

\renewcommand{\arraystretch}{1.7}
\begin{table}[h]
    \begin{center}
      \small
        \caption{Information integration result for the syllabus data}
        \label{tab:result_integration_syllabus}
        \scalebox{0.3}{
            \begin{tabular}{|c|c|c|c|c|c|c|} \hline
                Course Title & Instructor Name & Target Student & Target Year & Year/Term & Day/Period & Number of Credits \\ \hline
        \begin{tabular}{c}
        Instrumental Analysis, Adv. II
        \end{tabular} &
        \begin{tabular}{c}
        OOE KOUICHI \\
        Graduate School of Engineering Professor
        \end{tabular} &
        \begin{tabular}{c}
        Graduate Student
        \end{tabular} &
        \begin{tabular}{c}
        Master's student \\
        Doctoral student
        \end{tabular} &
        \begin{tabular}{c}
        Second semester
        \end{tabular} &
        \begin{tabular}{c}
        Thu. 4, 5
        \end{tabular} & 1\\ \hline
        \begin{tabular}{c}
        Advanced Studies: Educational of Clinical Psychologist II
        \end{tabular} &
        \begin{tabular}{c}
        NISHI MINAKO \\
        Graduate School of Education Associate Professor
        \end{tabular} &
        \begin{tabular}{c}
        Graduate student
        \end{tabular} &
        \begin{tabular}{c}
        Doctoral student
        \end{tabular} &
        \begin{tabular}{c}
        Second semester
        \end{tabular} &
        \begin{tabular}{c}
        Tue. 5
        \end{tabular} & 1 \\ \hline
        \begin{tabular}{c}
        Tax Law II
        \end{tabular} &
        \begin{tabular}{c}
        OKAMURA TADAO
        \end{tabular} &
        \begin{tabular}{c}
        Graduate student
        \end{tabular} &
        2, 3 &
        \begin{tabular}{c}
        Second semester
        \end{tabular} &
        \begin{tabular}{c}
        Fri. 5
        \end{tabular} &  \\ \hline
        \begin{tabular}{c}
        Advanced Materials Science \& Engineering in Industries II
        \end{tabular} &
        \begin{tabular}{c}
        TSUJI NOBUHIRO \\
        Graduate School of Engineering Professor
        \end{tabular} &
        \begin{tabular}{c}
        Graduate student
        \end{tabular} &
        \begin{tabular}{c}
        Master's student \\
        Doctoral student
        \end{tabular} &
        \begin{tabular}{c}
        Second semester
        \end{tabular} &
        \begin{tabular}{c}
        Tue. 4
        \end{tabular} & 2 \\ \hline
        \begin{tabular}{c}
        Advanced Reading on Clinical Psychology I
        \end{tabular} &
        \begin{tabular}{c}
        NATORI TAKUJI \\
        Part-time Lecturer
        \end{tabular} &
        \begin{tabular}{c}
        Graduate student
        \end{tabular} &
        \begin{tabular}{c}
        Master's student
        \end{tabular} &
        \begin{tabular}{c}
        First semester
        \end{tabular} &
        \begin{tabular}{c}
        Tue. 1
        \end{tabular} &  \\ \hline
        \begin{tabular}{c}
        Organic Chemistry in Food Science III
        \end{tabular} &
        \begin{tabular}{c}
        IRIE KAZUHIRO \\
        Graduate School of Agriculture Professor
        \end{tabular} &
        \begin{tabular}{c}
        Undergraduate student
        \end{tabular} &
        \begin{tabular}{c}
        2nd year students
        \end{tabular} &
        \begin{tabular}{c}
        Second semester
        \end{tabular} &
        \begin{tabular}{c}
        Tue. 3
        \end{tabular} & 2 \\ \hline
        \begin{tabular}{c}
        Advanced Environmental Biophysics
        \end{tabular} &
        \begin{tabular}{c}
          \begin{tabular}{c}
          SAKABE AYAKA \\
          Hakubi Center Assistant Professor
          \end{tabular} \\ \\
          \begin{tabular}{c}
          KOSUGI YOSHIKO \\
          Graduate School of Agriculture Professor
          \end{tabular}
        \end{tabular} &
        \begin{tabular}{c}
        Graduate student
        \end{tabular} &
        \begin{tabular}{c}
        \end{tabular} &
        \begin{tabular}{c}
        Intensive First semester
        \end{tabular} &
        \begin{tabular}{c}
        Intensive, First semester\\
        Scheduled for July 3, 27, 28, and 29
        \end{tabular}
        & 2 \\ \hline
        \begin{tabular}{c}
        Innovative Humano-habitability
        \end{tabular} &
        \begin{tabular}{c}
          \begin{tabular}{c}
          YANAGAWA AYA \\
          Research Institute for Sustainable Humanosphere Assistant Professor
          \end{tabular} \\ \\
          \begin{tabular}{c}
          HATA TOSHIMITSU \\
          Research Institute for Sustainable Humanosphere Lecturer
          \end{tabular} \\ \\
          \begin{tabular}{c}
          YOSHIMURA TSUYOSHI \\
          Research Institute for Sustainable Humanosphere Professor
          \end{tabular}
        \end{tabular} &
        \begin{tabular}{c}
        Graduate student
        \end{tabular} &
        \begin{tabular}{c}
        \end{tabular} &
        \begin{tabular}{c}
        Intensive, First semester
        \end{tabular} &
        \begin{tabular}{c}
        Intensive, 5/8 (Fri.), 5/15 (Fri.), 5/22 (Fri.)
        \end{tabular}
        & 2 \\ \hline
        \begin{tabular}{c}
        Exercises in Calligraphy and Copying B
        \end{tabular} &
        \begin{tabular}{c}
        HASEGAWA CHIHIRO \\
        Graduate School of Human and Environmental Studies Associate Professor
        \end{tabular} &
        \begin{tabular}{c}
        Undergraduate student
        \end{tabular} &
        \begin{tabular}{c}
        2nd--4th year students
        \end{tabular} &
        \begin{tabular}{c}
        Second semester
        \end{tabular} &
        \begin{tabular}{c}
        Wed. 2
        \end{tabular} & 2 \\ \hline
        \begin{tabular}{c}
        Energy and Information, Adv.
        \end{tabular} &
        \begin{tabular}{c}
        OOBAYASHI FUMIAKI \\
        Part-time Lecturer
        \end{tabular} &
        \begin{tabular}{c}
        Graduate student
        \end{tabular} &
        \begin{tabular}{c}
        Doctoral students
        \end{tabular} &
        \begin{tabular}{c}
        Intensive, First semester
        \end{tabular} &
        \begin{tabular}{c}
        Intensive
        \end{tabular} & 2 \\ \hline
        \begin{tabular}{c}
        Primary German A
        \end{tabular} &
        \begin{tabular}{c}
        KOMODA NAMI \\
        Part-time Lecturer
        \end{tabular} &
        \begin{tabular}{c}
        Undergraduate student
        \end{tabular} &
        \begin{tabular}{c}
        All students
        \end{tabular} &
        \begin{tabular}{c}
        First semester
        \end{tabular} &
        \begin{tabular}{c}
        Tue. 3
        \end{tabular} & 2 \\ \hline
         \end{tabular}
     }
    \end{center}
\end{table}
\renewcommand{\arraystretch}{1.0}

\newpage
\subsection{Ablation Experiments}
We conducted ablation studies to investigate the effect of adding root-to-leaf information transfer,
which is the opposite direction of the traditional Tree-LSTM, and the effect of HTML data augmentation introduced in this study.
In the HTML data augmentation, the order of any pair of rows and any pair of columns in the table was switched with a probability of 0.5.
The setting of the experiment is the same as the previous experiment on syllabus data,
and we compare the average values of all classes of $F_1$ measure of the traditional Tree-LSTM, our HTML-LSTM, and the HTML-LSTM with data augmentation.

The results are shown in the Table~\ref{tab:result_ablation_study}.
This result shows that the ability of information extraction can be improved by using not only the root-to-leaf direction but also the leaf-to-root direction.
We can also see that the data augmentation of HTML can further improve the accuracy of information extraction.

\renewcommand{\arraystretch}{1.2}
\begin{table}[t]
  \begin{center}
    \caption{Ablation study result}
    \label{tab:result_ablation_study}
    \begin{tabular}{cc} \thickhline
      Method & $F_1$ measure \\ \hline
      \begin{tabular}{c}
      Tree-LSTM~\cite{Tai2015}  \vspace{-2mm} \\
      {\tiny (Upward Tree-LSTM)}
      \end{tabular} & 0.8285 \\
      \begin{tabular}{c}
      HTML-LSTM \vspace{-2mm} \\
      {\tiny (Upward Tree-LSTM + Downward Tree-LSTM)}
      \end{tabular} & 0.8414 \\
      \begin{tabular}{c}
      HTML-LSTM \vspace{-2mm} \\
      {w/ HTML data augmentaion}
      \end{tabular} & 0.8575 \\
      \thickhline
    \end{tabular}
  \end{center}
\end{table}
\renewcommand{\arraystretch}{1.0}

\section{Conclusion}
In this paper, we proposed HTML-LSTM, a method for extracting and integrating required information from tables contained in multiple Web pages.
The method is an extension of Tree-LSTM, which is mainly used in the field of natural language processing and extracts words in texts attached to the leaves of DOM trees of HTML data in a bottom-up manner.
Our method treats DOM trees in a bottom-up manner and then a top-down manner to extract sequences of part-of-speeches and tags attached to nodes in the DOM trees.
We applied HTML-LSTM to a list of childcare facilities and syllabus data that are opened on the Web and confirmed that HTML-LSTM could extract information with $F_1$ measures of 0.96 and 0.86, respectively.

In the future, we would improve HTML-LSTM to extract information from fragments of HTML data other than tables.
Such fragments are also transformed into DOM trees.
For tables or lists, some special tags are prepared in HTML, but other fragments may not have such tags.
In order to overcome the problem, choosing good positive and negative examples would be important.
Also, modifying the HTML-LSTM algorithm would be needed.

\bibliographystyle{splncs03}
\bibliography{references}

\end{document}